# Eigen Solution and Thermodynamic Properties of Manning Rosen Plus Exponential Yukawa Potential

I. B. Okon[1], C. N. Isonguyo[1*], C. A. Onate[2], A. D. Antia[1], K. R. Purohit[3], E. E. Ekott[1], K.E. Essien[1], E.S. William[4] and N. E. Asuquo[1]

[1]*Theoretical Physics Group, Department of Physics, University of Uyo, Uyo, Nigeria*
[2]*Department of Physics, Kogi State University, Anyigba, Nigeria.*
[3]*Department of Physics, Sardar Vallabhbhai National Institute of Technology, Surat, Gujarat 395 007, India*
[4]*Theoretical Physics Group, Department of Physics, University of Calabar, Nigeria;*

The corresponding author email: *ceciliaisonguyo@uniuyo.edu.ng

**Abstract**
In this work, we obtained analytical bound state solution of the Schrödinger equation with Manning Rosen plus exponential Yukawa Potential using parametric Nikiforov-Uvarov method (NU). We obtained the normalized wave function in terms of Jacobi polynomial. The energy eigen equation was determined and presented in a compact form. The study also includes the computations of partition function and other thermodynamics properties such as vibrational mean energy ($\mu$), vibrational heat capacity (c), vibrational entropy (s) and vibrational free energy (F). Using a well design maple programme, we obtained numerical bound state energies for different quantum states with various screening parameters: $\alpha = 0.1, 0.2, 0.3, 0.4$ and $0.5$. The numerical results showed that the bound state energies increase with an increase in quantum state while the thermodynamic plots were in excellent agreement to work of existing literature.

## 1. Introduction

Thermodynamic properties of quantum system is one of the essential part in quantum physics. This has aroused the interest of many researchers especially its applications in physical sciences. The partition function (a function of temperature) is the central parameter for studying the thermodynamic properties of system, since other properties such as entropy, specific heat capacity, mean free energy depend on it. Thermodynamic properties can be achieved by using the solutions of the wave equations which consist of the eigenfunctions and eigenvalues that gives the conceptual understanding of quantum mechanical systems [1-5]. The relativistic wave equations are Dirac and Klein-Gordon equations while Schrödinger wave equation is a non-relativistic wave equation [6-11,26]. However, the determination of the approximate solutions of Schrödinger wave equation with various quantum potential models using different techniques such as : factorization method [12,13], Exact and proper quantization [14-16], Supersymmetry quantum mechanics (SUSY) [9,17-20],

Asymtotic Iteration method (AIM) [21], Nikiforov Uvarov method (NU) [8,22-25], Nikiforov Uvarov functional analysis method (NUFA) [42,43], amongst others have been studied over the years.

The application of a variety of quantum potential models to study the thermodynamic properties of systems have attracted many researchers in recent time. For instance, Edet et al. [28], used Poisson summation approach to study thermal properties of Deng-fan Eckart potential model. Ikot et al. [29], investigated the thermodynamic properties of diatomic molecules with general molecular potential. In ref [30], Ikot and his co-authors obtained the thermodynamic properties of Aharanov–Bohm (AB) and magnetic fields for screened Kratzer potential. Okorie et al. [31], calculated the thermodynamic properties of the modified Yukawa potential. Edet et al. [33], in one of the studies, carried out their research on thermal properties and magnetic susceptibility of Hellmann potential in the presence Aharonov–Bohm (AB) flux and magnetic fields at zero and finite temperatures. Okon et al. [23] combined Mobius Square and Screened Kratzer Potential to obtained thermodynamic properties and bound State Solutions to Schrödinger equation. Other potential models such as Hellmann potential [33], Mixed Hyperbolic Poschl-Teller Potential [34], Kratzer Plus generalized Morse Potential [35], Hyperbolic Hulthen plus hyperbolic exponential inversely quadratic potential [27] were also studied.

The breakthrough in the work of these researchers have motivated us to consider the bound state solutions of Schrödinger equation and its application to partition function and other thermodynamic properties with Manning Rosen Plus exponential Yukawa Potential using parametric Nikiforov-Uvarov method. The Manning Rosen Plus exponential Yukawa Potential is one of the short-range potentials. These potentials (especially the Manning-Rosen potential) give an excellent description of the interaction between diatomic molecules. The potentials find applications in several branches of physics such as high energy physics, molecular physics and others [7, 36-41]. The Manning Rosen Plus exponential Yukawa Potential is given as [22-23, 58]

$$V(r) = -\left(\frac{A_1 e^{-\alpha r} + A_2 e^{-2\alpha r}}{(1-e^{-\alpha r})^2}\right) - \frac{A_3 e^{-\alpha r}}{r} \qquad (1)$$

where $A_1$, $A_2$ and $A_3$ are dimensionless parameters, r is the internuclear separation and $\alpha$ is the screening parameter or range of the potential. This useful potential only received in ref. [58] under the Klein-Gordon equation. The present work will determine the nonrelativistic solutions of this potential and extends its application to thermodynamic properties.

## 2. PARAMETRIC NIKIFOROV – UVAROV METHOD

The parametric formalization of NU involves reducing the second order linear differential equation to a generalized equation of hyper-geometric-type. This method provides exact solutions in terms of special orthogonal functions as well as the corresponding energy equation. With appropriate coordinate transformation, $s = s(x)$, this equation can be written as [44-47].

$$\psi''(s) + \frac{\bar{\tau}(s)}{\sigma(s)}\psi^1{}_{(s)} + \frac{\tilde{\sigma}(s)}{\sigma^2(s)}\psi(s) = 0 \qquad (2)$$

where $\bar{\tau}(s)$ is a polynomial of degree one, $\sigma(s)$ and $\tilde{\sigma}(s)$ are polynomials of at most degree two. Then the parametric NU differential equation is in the form [].

$$\psi''(s) + \frac{c_1 - c_2 s}{s(1 - c_3 s)}\psi'(s) + \frac{1}{s^2(1 - c_3 s)^2}[-\xi_1 s^2 + \xi_2 s - \xi_3]\psi(s) = 0 \qquad (3)$$

The parametric constants are obtained as follows

$$\begin{bmatrix} c_1 = c_2 = c_3 = 1, \ c_4 = \frac{1}{2}(1 - c_1), \ c_5 = \frac{1}{2}(c_2 - 2c_3), \\ c_6 = c_5^2 + \xi_1, \ c_7 = 2c_4 c_5 - \xi_2, \ c_8 = c_4^2 + \xi_3, \\ c_8 = c_4^2 + \xi_3, \ c_9 = c_3 c_7 + c_3^2 c_8 + c_6, \ c_{10} = c_1 + 2c_4 + 2\sqrt{c_8}, \\ c_{11} = c_2 - 2c_5 + 2(\sqrt{c_9} + c_3\sqrt{c_8}), c_{12} = c_4 + \sqrt{c_8} \\ c_{13} = c_5 - (\sqrt{c_9} + c_3\sqrt{c_8}) \end{bmatrix} \qquad (4)$$

The condition for energy equation is written in the form

$$c_2 n - (2n + 1)c_5 + (2n + 1)(\sqrt{c_9} + c_3\sqrt{c_8}) + n(n - 1)c_3 + c_7 + 2c_3 c_8 + 2\sqrt{c_8 c_9} = 0, \qquad (5)$$

while corresponding wave function is given by

$$\psi(s) = \phi(s)\chi_n(s) = N_n s^{-c_{12} - \frac{c_{13}}{c_3}} P_n^{\left(c_{10}-1, \frac{c_{11}}{c_3} - c_{10}-1\right)}(1 - 2c_3 s) \qquad (6)$$

## 3. Radial Solution of Schrödinger Wave Equation With the Potential Model

The Schrödinger wave equation for an arbitrary external potential V(r) in spherical coordinate is written as [48-57] as

$$\frac{d^2 R(r)}{dr^2} + \frac{2\mu}{\hbar^2}\left\{(E_{nl} - V(r)) - \frac{\hbar^2 l(l+1)}{2\mu r^2}\right\}\psi(r) = 0 \qquad (7)$$

where $E_{nl}$ is the exact bound state energy eigenvalues, $\psi(r)$ is the wave function, μ represent the reduced mass. n and $l$ are known as the quantum number and rotation quantum number. On substituting equation (1) into equation (7), the radial part of the Schrödinger equation for the Manning Rosen Plus exponential Yukawa Potential becomes

$$\frac{d^2R(r)}{dr^2} + \left\{\frac{2\mu}{\hbar^2}\left[E_{nl} + \left(\frac{A_1 e^{-\alpha r} + A_2 e^{-2\alpha r}}{(1-e^{-\alpha r})^2}\right) + \frac{A_3 e^{-\alpha r}}{r}\right] - \frac{l(l+1)}{r^2}\right\}\psi(r) = 0. \tag{8}$$

To deal with the present of the centrifugal barrier in equation (8) above, the Greene–Aldrich approximation scheme is employed

$$\frac{1}{r^2} = \frac{\alpha^2}{(1-e^{-\alpha r})^2} \Rightarrow \frac{1}{r} = \frac{\alpha}{(1-e^{-\alpha r})} \tag{9}$$

On putting the transformation $s = e^{-\alpha r}$ and applying the Green-Aldrich approximation in equation (9), equation (8) yields

$$\frac{d^2R}{ds^2} + \frac{(1-s)}{s(1-s)}\frac{dR(s)}{ds} + \frac{1}{s^2(1-s)^2}\{-(\xi^2 - x_2 + x_3)s^2 + (2\xi^2 + x_1 + x_3)s - (\xi^2 + l(l+1))\}\psi(s), = 0 \tag{10}$$

where

$$\xi^2 = \frac{-2\mu E_{nl}}{\hbar^2 \alpha^2}, \quad x_1 = \frac{2\mu A_1}{\hbar^2 \alpha^2}, x_2 = \frac{2\mu A_2}{\hbar^2 \alpha^2}, x_3 = \frac{2\mu A_3}{\hbar^2 \alpha} \tag{11}$$

Comparing (10) to (3), we obtained the following parameters

$$\begin{bmatrix} \chi_1 = \xi^2 - x_2 + x_3, \chi_2 = 2\xi^2 + x_1 + x_3, \chi_3 = \xi^2 + l(l+1), \\ c_1 = c_2 = c_3 = 1, c_4 = 0, c_5 = -\frac{1}{2}, c_6 = \frac{1}{4} + \xi^2 - x_2 + x_3, \\ c_7 = -2\xi^2 - x_1 - x_3, c_8 = \xi^2 l(l+1), c_9 = \frac{1}{4} + l(l+1) - x_1 - x_2, \\ c_{10} = 1 + 2\sqrt{\xi^2 + l(l+1)}, c_{12} = \sqrt{\xi^2 + l(l+1)}, \\ c_{11} = 2 + 2\left[\sqrt{\frac{1}{4} + l(l+1) - x_1 - x_2} + \sqrt{\xi^2 + l(l+1)}\right], \\ c_{13} = -\frac{1}{2} - \left(\sqrt{\frac{1}{4} + l(l+1) - x_1 - x_2} + \sqrt{\xi^2 + l(l+1)}\right). \end{bmatrix} \tag{12}$$

Using the condition for the energy eigen equation (5) and substituting all the parametric constants in equation (12) and simplify, the total energy eigen equation is then given as

$$E_{nl} = \frac{-\hbar^2 \alpha^2}{2\mu}\left\{\frac{\left(n^2 + n + \frac{1}{2}\right)\left(n + \frac{1}{2}\right)\sqrt{1 + 4l(l+1) - \frac{8\mu A_1}{\hbar^2 \alpha^2} - \frac{8\mu A_2}{\hbar^2 \alpha^2} - \frac{2\mu A_1}{\hbar^2 \alpha^2} - \frac{2\mu A_3}{\hbar^2 \alpha^2} + 2l(l+1)}}{2n + 1 + \sqrt{1 + 4l(l+1) - \frac{8\mu A_1}{\hbar^2 \alpha^2} - \frac{8\mu A_2}{\hbar^2 \alpha^2}}}\right\}^2 + \frac{\hbar^2 \alpha^2 l(l+1)}{2\mu}. \tag{13}$$

Equation (13) can be presented in closed and compact from as

$$E_{nl} = Q_1 - Q_2\left\{(n + \delta) + \frac{Q_3}{(n+\delta)}\right\}^2 \tag{14}$$

where

$$\begin{cases} Q_1 = \frac{\hbar^2\alpha^2 l(l+1)}{2\mu}, Q_2 = \frac{\hbar^2\alpha^2}{8\mu}, Q_3 = \frac{2\mu A_2}{\hbar^2\alpha^2} - \frac{2\mu A_3}{\hbar^2\alpha} + l(l+1), \\ \delta = \frac{1}{2} + \frac{1}{2}\sqrt{1+4l(l+1) - \frac{8\mu A_1}{\hbar^2\alpha^2} - \frac{8\mu A_2}{\hbar^2\alpha^2}} \end{cases} \qquad (15)$$

The radial wave function in equation (6) then reduces to

$$\psi(r) = N_n(e^{-\alpha r})^\beta (1-e^{-\alpha r})^\zeta P_n^{(2\beta,\ 2\zeta-1)}(1-2e^{-\alpha r}) \qquad (16)$$

where

$$\beta = \sqrt{\frac{-2\mu E_n}{\hbar^2\alpha^2} + l(l+1)}, \quad \zeta = \frac{1}{2}\sqrt{\frac{1}{4} + l(l+1) - \frac{2\mu A_1}{\hbar^2\alpha^2} - \frac{2\mu A_2}{\hbar^2\alpha^2}} \qquad (17)$$

### 3.1 Special Cases

The interacting potential has two special cases. The potential reduces to Manning Rosen and exponential Yukuwa potentials

**(a) Manning Rosen Potential**

Putting $A_3 = 0$ into the (1), the potential reduces to Manning Rosen potential of the form

$$V(r) = -\left(\frac{A_1 e^{\alpha r} + A_2 e^{-2\alpha r}}{(1-e^{-\alpha r})^2}\right) \qquad (18)$$

Similar, putting $A_3 = 0$ in the energy equation (13) gives the energy of Manning Rosen potential as

$$E = \frac{\hbar^2\alpha^2 l(l+1)}{2\mu} - \frac{\hbar^2\alpha^2}{8\mu}\left[\frac{\left[n+\frac{1}{2}+\frac{1}{2}\sqrt{1-\frac{8\mu c_1}{\hbar^2\alpha^2}-\frac{8\mu c_2}{\hbar^2\alpha^2}+4l(l+1)}\right]^2 + \frac{2\mu c_1}{\hbar^2\alpha^2}+l(l+1)}{\left[n+\frac{1}{2}+\frac{1}{2}\sqrt{1-\frac{8\mu c_1}{\hbar^2\alpha^2}-\frac{8\mu c_2}{\hbar^2\alpha^2}+4l(l+1)}\right]}\right]^2 \qquad (19)$$

**b. Exponential Yukawa potential**

On putting $A_1 = A_2 = 0$, in (1), the potential reduces to exponential Yukawa potential

$$V(r) = -\frac{A_3 e^{-\alpha r}}{r} \qquad (20)$$

Then energy equation (13) reduces to the energy equation for the Yukawa potential as

$$E = \frac{\hbar^2\alpha^2 l(l+1)}{2\mu} - \frac{\hbar^2\alpha^2}{2\mu}\left[\frac{\left(n^2+n+\frac{1}{2}\right)+\left(n+\frac{1}{2}\right)\sqrt{1-4l(l+1)} - \frac{2\mu A_3}{\hbar^2\alpha}+l(l+1)}{2n+1\sqrt{1-4l(l+1)}}\right]^2 \qquad (21)$$

## 4. Thermomagnetic Properties of Manning Rosen Plus exponential Yukawa Potential

The thermodynamic properties of quantum systems can be obtained from the exact partition function given as

$$Z(\beta) = \sum_{n=0}^{\lambda} e^{-\beta E_n}, \qquad (22)$$

Where, $\lambda$ an upper bound of the vibrational quantum number obtain from the numerical solution of $\frac{dE_n}{dn} = 0$, $\beta = \frac{1}{kT}$, K and T are Boltzmann constant and absolute temperature respectively. In classical limit, the summation in equation (22) can be replaced with the integral:

$$Z(\beta) = \int_0^{\lambda} e^{-\beta E_n} dn \qquad (23)$$

By inserting equation (14) into (23) using Maple 10.0 version, the partition function becomes

$$Z(\beta) = e^{\beta(2Q_2Q_3 - Q_1)} \int_0^{\lambda} e^{\beta\left(Q_2\rho^2 + \frac{Q_2Q_3^2}{\rho^2}\right)} d\rho \qquad (24)$$

Other thermodynamic properties are obtained with the help of equation (24) and (25) as follows:

(a) Vibrational mean energy

$$U(\beta) = -\frac{\partial \ln Z(\beta)}{\partial \beta} \qquad (26)$$

(b) Vibrational entropy

$$S(\beta) = K \ln Z(\beta) - K\beta \frac{\partial \ln Z(\beta)}{\partial \beta} \qquad (27)$$

(c) Vibrational Free Energy

$$F(\beta) = -\frac{1}{\beta} \ln Z(\beta) \qquad (28)$$

(d) Vibrational Heat Capacity

$$C(\beta) = K\beta^2 \left(\frac{\partial^2 \ln Z(\beta)}{\partial \beta^2}\right) \qquad (29)$$

## 5. Numerical Solutions And Discussion

**Table 1:** Numerical Bound State Solution for the Proposed Potential for $\alpha = 0.1$

| n | l | $E_{nl}$ | l | $E_{nl}$ | l | $E_{nl}$ | l | $E_{nl}$ |
|---|---|----------|---|----------|---|----------|---|----------|
| 0 | 0 | 0.087559 | 1 | 0. 094256 | 2 | 0.110987 | 3 | 0.137753 |
| 1 | 0 | 0.084874 | 1 | 0.091542 | 2 | 0.108243 | 3 | 0.134976 |
| 2 | 0 | 0.079538 | 1 | 0.086156 | 2 | 0.102801 | 3 | 0.129474 |
| 3 | 0 | 0.071607 | 1 | 0.078161 | 2 | 0.094737 | 3 | 0.121334 |
| 4 | 0 | 0.061142 | 1 | 0.067628 | 2 | 0.084129 | 3 | 0.110647 |
| 5 | 0 | 0.048191 | 1 | 0.054612 | 2 | 0.071044 | 3 | 0.097486 |

**Table 2:** Numerical Bound State Solution for the Proposed Potential for ∝=0.2

| n | l | $E_{nl}$ | l | $E_{nl}$ | l | $E_{nl}$ | l | $E_{nl}$ |
|---|---|---|---|---|---|---|---|---|
| 0 | 0 | 0.094264 | 1 | 0. 120764 | 2 | 0.187817 | 3 | 0.295496 |
| 1 | 0 | 0.083889 | 1 | 0.110044 | 2 | 0.176600 | 3 | 0.283593 |
| 2 | 0 | 0.063366 | 1 | 0.089056 | 2 | 0.154952 | 3 | 0.261059 |
| 3 | 0 | 0.032911 | 1 | 0.058210 | 2 | 0.123571 | 3 | 0.228979 |
| 4 | 0 | -0.007400 | 1 | 0.017645 | 2 | 0.082669 | 3 | 0.187653 |
| 5 | 0 | -0.057582 | 1 | -0.032678 | 2 | 0.032166 | 3 | 0.136934 |

**Table 3:** Numerical Bound State Solution for the Proposed Potential for ∝=0.3

| n | l | $E_{nl}$ | l | $E_{nl}$ | l | $E_{nl}$ | l | $E_{nl}$ |
|---|---|---|---|---|---|---|---|---|
| 0 | 0 | 0.100139 | 1 | 0. 159398 | 2 | 0.311614 | 3 | 0.557808 |
| 1 | 0 | 0.077320 | 1 | 0.135208 | 2 | 0.284403 | 3 | 0.524963 |
| 2 | 0 | 0.0320150 | 1 | 0.088683 | 2 | 0.235449 | 3 | 0.472002 |
| 3 | 0 | -0.035648 | 1 | 0.020429 | 2 | 0.166161 | 3 | 0.401279 |
| 4 | 0 | -0.125711 | 1 | -0.069835 | 2 | 0.075613 | 3 | 0.310458 |
| 5 | 0 | -0.238227 | 1 | -0.182384 | 2 | -0.036938 | 3 | 0.198002 |

**Table 4:** Numerical Bound State Solution for the Proposed Potential for ∝=0.4

| n | l | $E_{nl}$ | l | $E_{nl}$ | l | $E_{nl}$ | l | $E_{nl}$ |
|---|---|---|---|---|---|---|---|---|
| 0 | 0 | 0.105184 | 1 | 0. 210385 | 2 | 0.486021 | 3 | 0.940000 |
| 1 | 0 | 0.065069 | 1 | 0.166635 | 2 | 0.429536 | 3 | 0.850748 |
| 2 | 0 | -0.015002 | 1 | 0.084691 | 2 | 0.342832 | 3 | 0.756976 |
| 3 | 0 | -0.135023 | 1 | -0.035761 | 2 | 0.221837 | 3 | 0.636598 |
| 4 | 0 | -0.295026 | 1 | -0.195747 | 2 | 0.062202 | 3 | 0.478259 |
| 5 | 0 | -0.495023 | 1 | -0.395635 | 2 | -0.137266 | 3 | 0.279793 |

**Table 5:** Numerical Bound State Solution for the Proposed Potential for ∝=0.5

| n | l | $E_{nl}$ | l | $E_{nl}$ | l | $E_{nl}$ | l | $E_{nl}$ |
|---|---|---|---|---|---|---|---|---|
| 0 | 0 | 0.109375 | 1 | 0. 274306 | 2 | 0.718750 | 3 | 1.468750 |
| 1 | 0 | 0.046875 | 1 | 0.203750 | 2 | 0.605469 | 3 | 1.218750 |
| 2 | 0 | -0.078125 | 1 | 0.076775 | 2 | 0.475586 | 3 | 1.109056 |
| 3 | 0 | -0.265625 | 1 | -0.110694 | 2 | 0.290179 | 3 | 0.933987 |
| 4 | 0 | -0.515625 | 1 | -0.360409 | 2 | 0.042097 | 3 | 0.690689 |
| 5 | 0 | -0.828125 | 1 | -0.672664 | 2 | -0.269104 | 3 | 0.381991 |

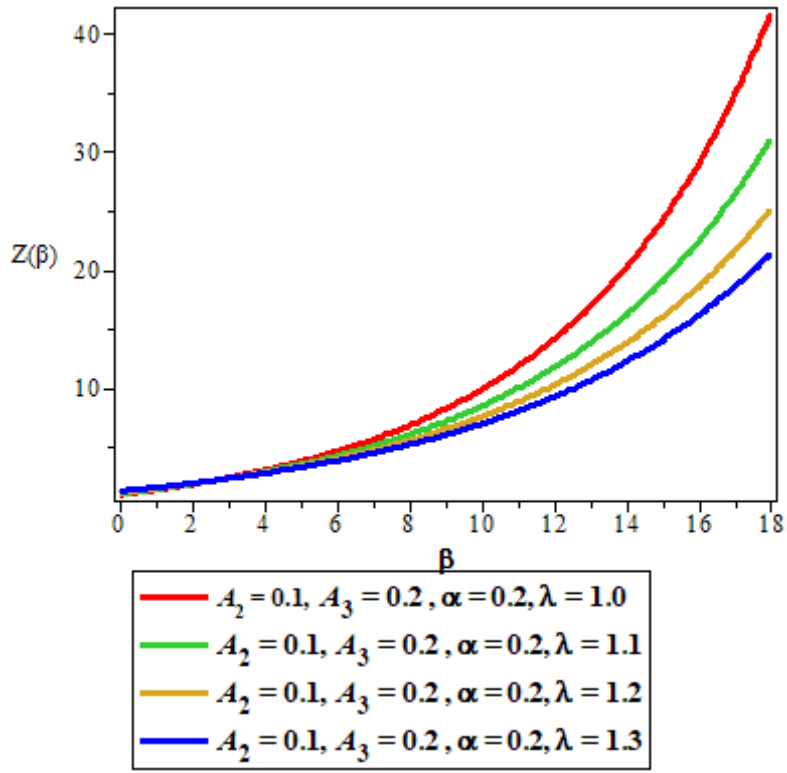

Figure 1: Variation of partition function with respect to inverse temperature parameter (β)

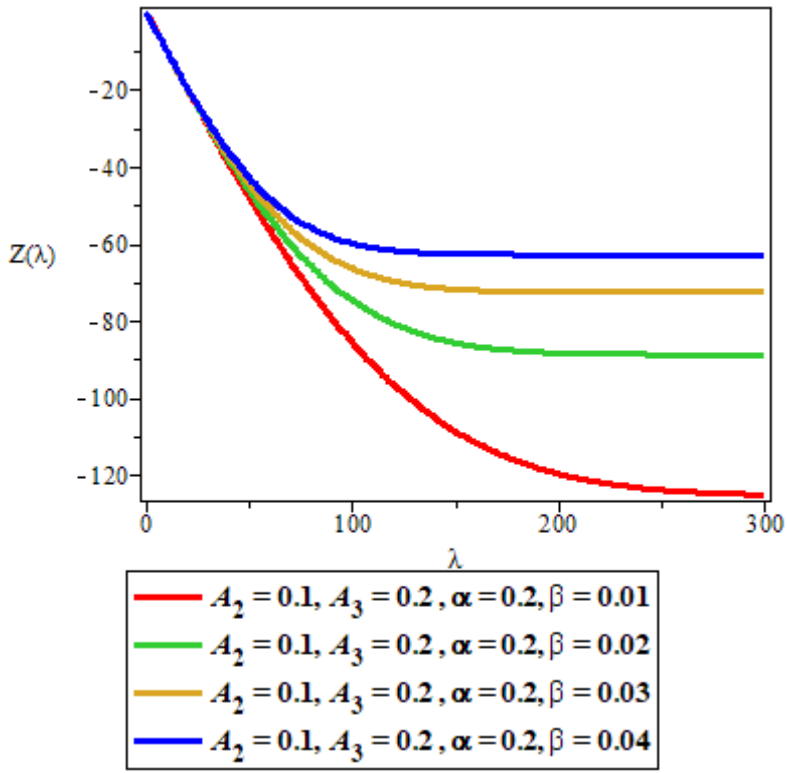

Figure 2: Variation of partition function with respect to maximum vibrational quantum number (λ)

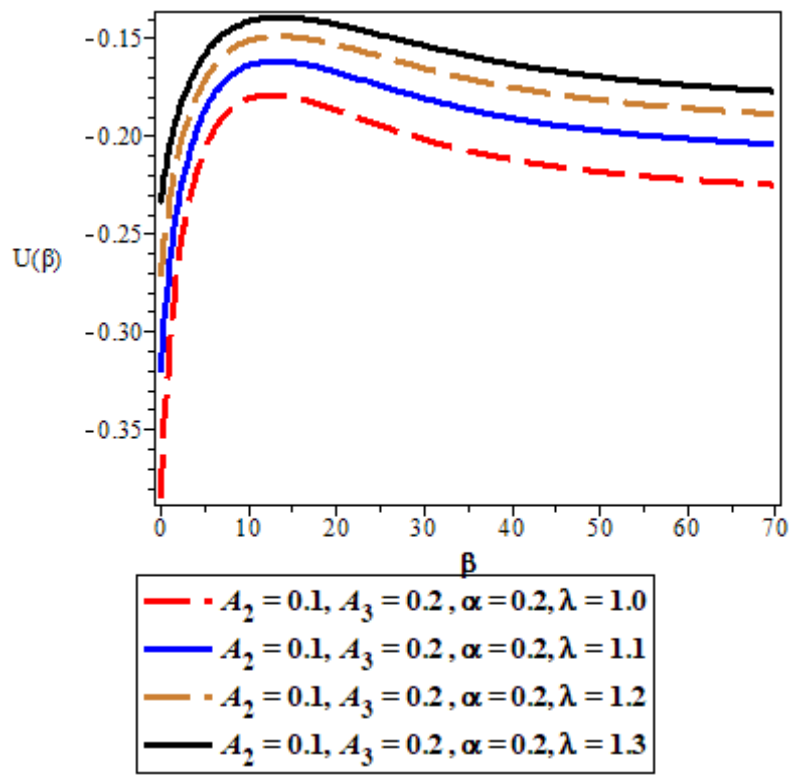

Figure 3: Variation of vibrational mean energy with respect to inverse temperature parameter (β)

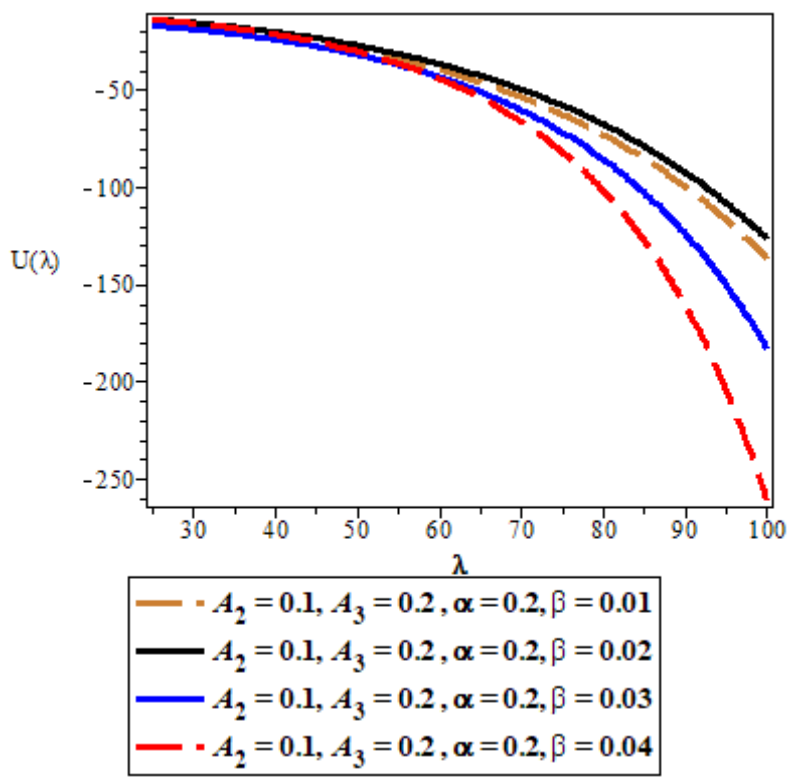

Figure 4: Variation of vibrational mean energy with respect to maximum quantum number (λ)

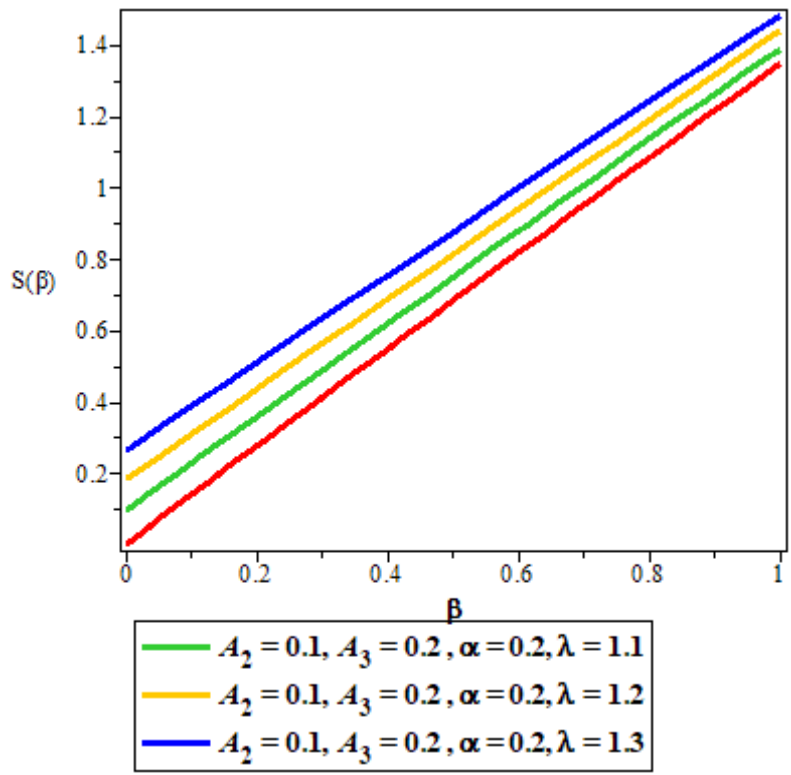

Figure 5: Variation of vibrational entropy with respect to inverse temperature parameter ($\beta$)

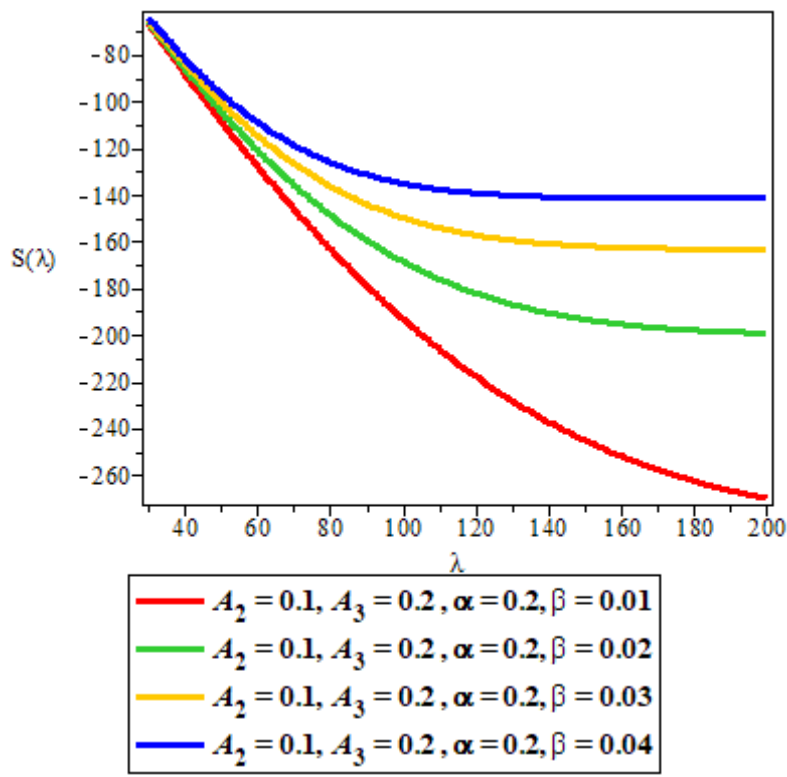

Figure 6: Variation of vibrational entropy with respect to maximum quantum number ($\lambda$)

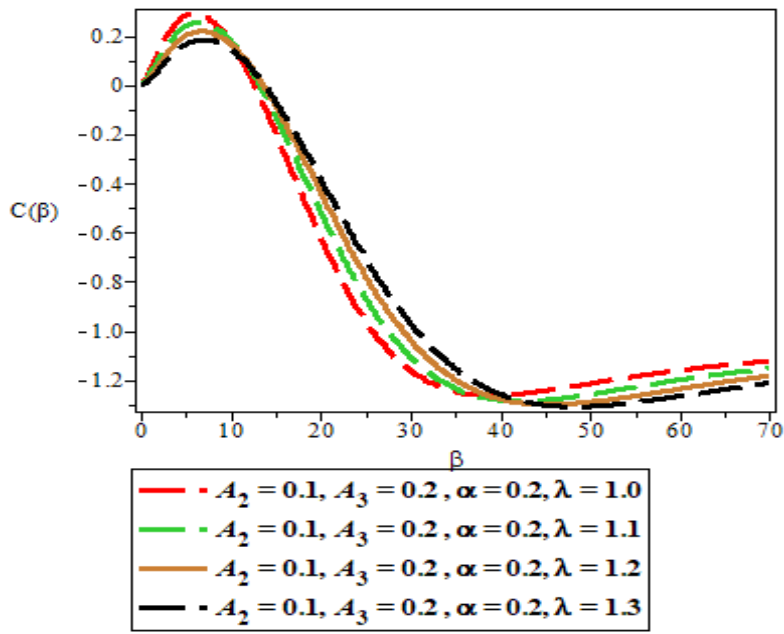

Figure 7: Variation of heat capacity with respect to inverse temperature parameter ($\beta$)

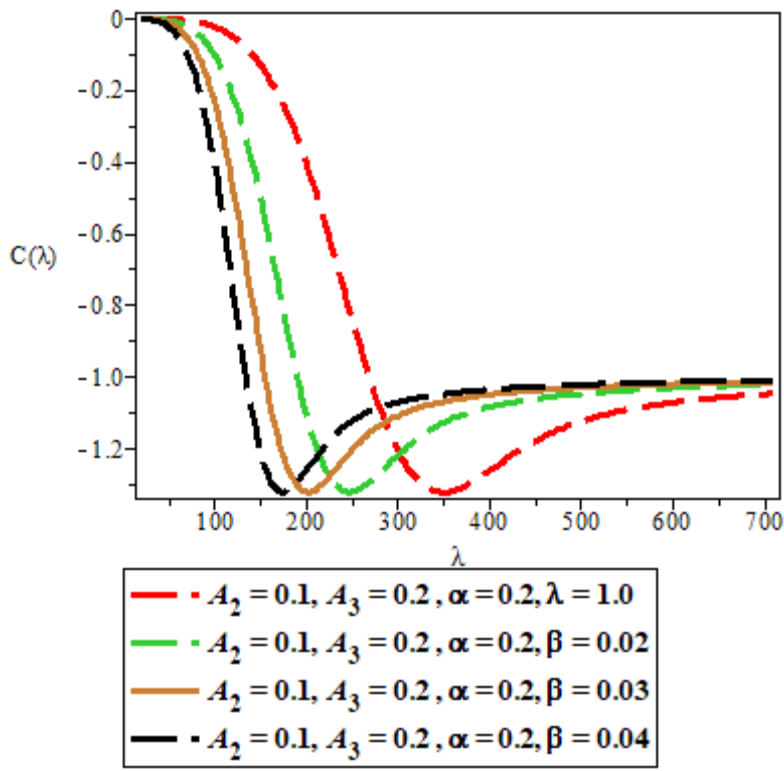

Figure 8: Variation of heat capacity with respect to inverse temperature parameter ($\lambda$)

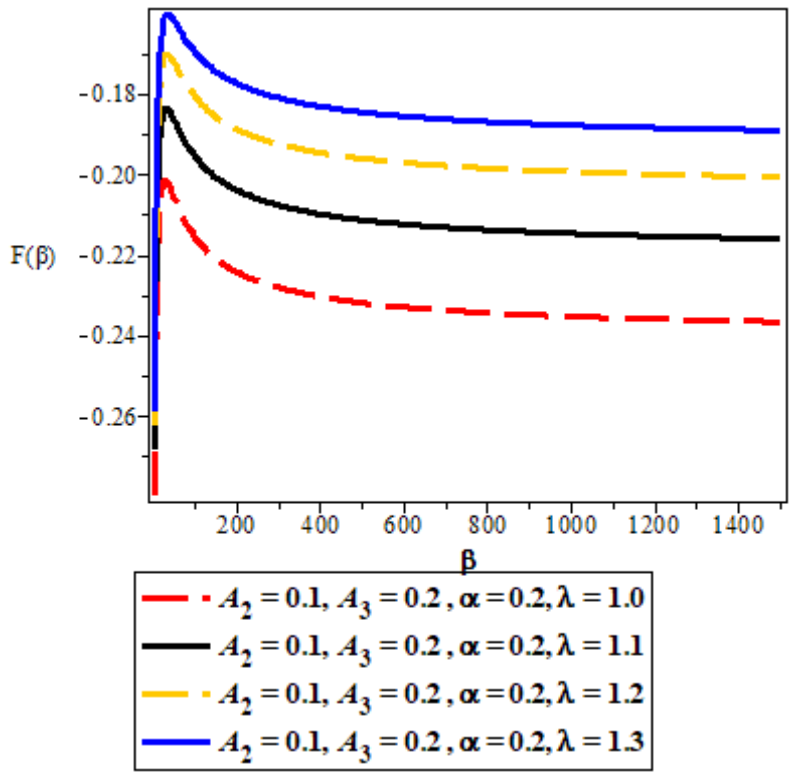

Figure 9: Variation of free energy with respect to inverse temperature parameter (β)

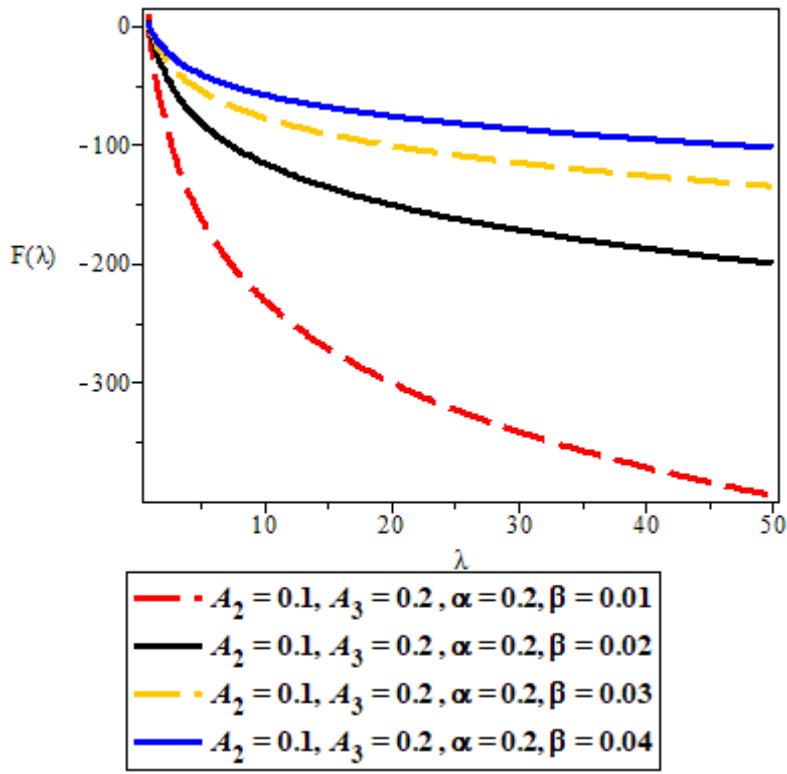

Figure 10: Variation of free energy with respect to maximum quantum number (λ)

## 6 Discussion

In table 1 the numerical bound state energies decrease with an increase in quantum state for various orbital angular quantum number (l) for $\alpha = 0.1$. The same trend is observed in tables 2, 3, 4 and 5. Some values obtained in tables 2, 3, 4 and 5 are negative which authenticate bound state conditions for $\alpha = 0.2, 0.3, 0.4$ and $0.5$ respectively. It can also be observed that the numerical bound state solution increases with an increase in orbital angular quantum number ($l$). The numerical bound state solution for a particular value of screening parameter ($\alpha$) has a distinct value to ascertain the boundedness of individual localized particle within the potential well. In figure 1, the partition function increases exponentially from the origin before splitting into different spectral curves in an increasing value of inverse temperature parameter ($\beta$). In figure 2, the partition function starts from the positive vertical axis and increase exponentially before splitting into different spectral graphs with respect to maximum vibrational quantum number($\lambda$). In figure 3, the vibrational mean energy increases monotonically from the origin before splitting into a unique quantized curves which increases with increasing value of inverse temperature parameter.

In figure 4, the vibrational mean energy is a non-linear curve that increases exponentially from the negative vertical-axis and increases with increasing value of maximum vibrational quantum number. In figure 5, the graph of the vibrational entropy is a linear graph that has the same explanation as figure 2. In figure 6, the vibrational heat capacity is a sinusoidal graph with various turning point. It has maximum turning at $\beta = 10\text{k}^{-1}$ and minimum turning point at $\beta = 40\text{k}^{-1}$

Figure 8 is sinusoidal curve that has a distinctive minimum turning points at $\lambda = 180, 200, 250 \text{ and } 350$ respectively before converging at $\lambda = 700$. In figure 9, the vibrational free energy F($\beta$) increases monotonically from the origin with a sharp peak at $\beta = 10k^{-1}$ before increasing linearly in a unique and quantized manner. In figure 10, the vibrational free energy $F(\beta)$ increases exponentially from the negative y-axis with large and conspicuous spacing.

## 7. Conclusion

In this work, we considered an analytical bound state solution of Schrŏdinger equation with Manning Rosen plus exponential Yukawa potential using parametric Nikiforov-Uvarov method with the help of standard

Greene-Aldrich approximation to centrifugal term. We obtained energy eigen equation and the normalized radial wave function expressed in terms of confluent hyper-geometric function of Jacobi polynomial. By adjusting some physical constant parameters, the potential reduces to both Manning-Rosen and exponential Yukawa as special cases. The numerical bound state solutions were obtained for various values of the screening parameter ($\alpha = 0.1, 0.2, 0.3, 0.4\ and\ 0.5$). The energy eigen equation were presented in a closed form and applied to study partition function (Z) and other thermodynamic properties. The trend of the thermodynamic plots were in excellent agreement to the published work to which ascertain the accuracy of the analytical calculation of the present study. The current research work has application in chemical and physical sciences as well as particle physics.